\documentclass[sigconf,nonacm,authorversion]{acmart}
\acmConference[CAIN 2024]{3rd International Conference on AI Engineering — Software Engineering for AI}{April 2024}{Lisbon, Portugal}

\AtBeginDocument{%
  }
\usepackage{tikz}
\usetikzlibrary{arrows.meta,shapes,positioning,shadows,trees,shapes.geometric}
\usepackage{paralist}
\usepackage{graphicx}
\usepackage{xcolor}
\usepackage{tcolorbox}
\usepackage{booktabs}
\usepackage{framed}
\usepackage{forest}
\usepackage{pgfplots}
\pgfplotsset{compat=1.17}
\usepackage[linesnumbered, ruled, vlined]{algorithm2e}
\SetAlFnt{\small}
\SetAlCapFnt{\small}
\SetAlCapNameFnt{\small}
\SetKwRepeat{Do}{do}{while}

\usepackage{xcolor}
\usepackage{listings}
\definecolor{codegreen}{rgb}{0,0.6,0}
\definecolor{codegray}{rgb}{0.5,0.5,0.5}
\definecolor{codepurple}{rgb}{0.58,0,0.82}
\definecolor{backcolour}{rgb}{0.95,0.95,0.92}
\lstdefinestyle{mystyle}{
    backgroundcolor=\color{backcolour},   
    commentstyle=\color{codegreen},
    keywordstyle=\color{magenta},
    numberstyle=\tiny\color{codegray},
    stringstyle=\color{codepurple},
    basicstyle=\ttfamily\small,
    breakatwhitespace=false,         
    breaklines=true,                 
    captionpos=b,                    
    keepspaces=true,                 
    numbers=none,                    
    showspaces=false,                
    showstringspaces=false,
    showtabs=false,                  
    tabsize=2,
    frame=single,
    framesep=2pt,
    boxpos=c
}
\begin{document}

\title{Mutation-based Consistency Testing for Evaluating the Code Understanding Capability of LLMs}

\author{Ziyu Li}
\email{zli311@sheffield.ac.uk}
\orcid{0009-0004-9821-3174}
\affiliation{%
  \institution{University of Sheffield}
  \city{Sheffield}
  \country{UK}
}

\author{Donghwan Shin}
\email{d.shin@sheffield.ac.uk}
\orcid{0000-0002-0840-6449}
\affiliation{%
  \institution{University of Sheffield}
  \city{Sheffield}
  \country{UK}
}

\renewcommand{\shortauthors}{Li and Shin}

\begin{abstract}
Large Language Models (LLMs) have shown remarkable capabilities in processing both natural and programming languages, which have enabled various applications in software engineering, such as requirement engineering, code generation, and software testing. However, existing code generation benchmarks do not necessarily assess the code understanding performance of LLMs, especially for the subtle inconsistencies that may arise between code and its semantics described in natural language.

In this paper, we propose a novel method to systematically assess the code understanding performance of LLMs, particularly focusing on subtle differences between code and its descriptions, by introducing code mutations to existing code generation datasets. Code mutations are small changes that alter the semantics of the original code, creating a mismatch with the natural language description. We apply different types of code mutations, such as operator replacement and statement deletion, to generate inconsistent code-description pairs. We then use these pairs to test the ability of LLMs to correctly detect the inconsistencies.

We propose a new LLM testing method, called Mutation-based Consistency Testing (MCT), and conduct a case study on the two popular LLMs, GPT-3.5 and GPT-4, using the state-of-the-art code generation benchmark, HumanEval-X, which consists of six programming languages (Python, C++, Java, Go, JavaScript, and Rust). We compare the performance of the LLMs across different types of code mutations and programming languages and analyze the results. We find that the LLMs show significant variation in their code understanding performance and that they have different strengths and weaknesses depending on the mutation type and language. 
We further explain conditions under which the LLMs result in correct answers using input characteristics (e.g., number of tokens) and investigate to what extent the test results can be improved using one-shot prompts (i.e., providing an additional example). 
Our MCT method and the case study results provide valuable implications for future research and development of LLM-based software engineering. 
\end{abstract}



\maketitle

\section{Introduction}\label{sec:intro}
With the recent advances in Large Language Models (LLMs), which can deal with even code comprehension and generation, LLM-based Software Engineering (SE)~\cite{fan2023large,hou2023large} has emerged to address various software engineering problems, ranging from requirement engineering to software testing. 
Considering the capabilities of LLMs in processing both natural languages and programming languages, it is not surprising that the number of LLM-based SE papers has grown dramatically over the last few years (from 11 papers in 2021 to 160 papers in the first half of 2023)~\cite{hou2023large}. 

LLMs have also been actively used in software development practice. 
LLM-based code completion tools, such as Copilot, have been integrated into popular IDEs, such as Visual Studio Code and PyCharm. 
More LLM-based tools for code completion, code generation, and code interpretation are coming, such as TabNine~\cite{tabnine}, Codex~\cite{finnie2022robots,prenner2022can}, and Figstack~\cite{figstack}. 
A recent survey from GitHub~\cite{github-blog} showed that 92\% of U.S.-based developers in large companies admit using AI tools, and 70\% say it to be useful. 
Such increasing attention to LLM-based SE from both researchers and practitioners calls for more rigorous and systematic testing of the programming capabilities of LLMs. 

Notable examples in the early stage of LLM testing are code generation benchmarks, such as HumanEval~\cite{chen2021evaluating} and HumanEval-X~\cite{zheng2023codegeex}.
They evaluate whether LLMs can generate a correct program (e.g., function) from a natural language description (e.g., docstring), focusing on the code writing rather than the code reading capability of LLMs.
Recently, \citet{ma2023scope} investigated the code syntax and semantics understanding capability of LLMs based on Abstract Syntax Tree 
(AST) and Control Flow Graph (CFG). 
However, correctly understanding code structures does not necessarily mean that the semantics of programs are correctly understood. 
For instance, two simple code fragments `\texttt{a=x+y}' and `\texttt{a=x-y}' have the same AST structure but totally different semantics. 

In this paper, we aim to systematically evaluate the programming capability of LLMs, particularly in terms of identifying subtle inconsistencies between code and its corresponding semantics (given as natural language descriptions). 
To achieve this, we present a new LLM testing method, called \textit{Mutation-based Consistency Testing} (MCT) for LLMs. 
The key idea behind MCT is to apply code mutations to the correct programs provided in code generation benchmarks to create subtle inconsistencies between the code and its description.
Using programming mutation operators widely studied in mutation analysis~\cite{petrovic2018state, petrovic2021practical, hamimoune2016mutation}, we can systematically generate various semantic inconsistencies. 
Furthermore, by analyzing the incorrect answers of LLMs based on mutation operators applied, we can provide more in-depth analysis results on the strengths and weaknesses of LLMs in understanding the code semantics. 

We conducted a case study on the two most popular LLMs, GPT-3.5 and GPT-4, based on a state-of-the-art code generation benchmark, HumanEval-X~\cite{zheng2023codegeex}, which consists of six programming languages (Python, C++, Java, Go, JavaScript, and Rust). We investigated how different mutation operators and programming languages influence the performance of these LLMs. Additionally, we examined whether the LLMs' understanding varies based on the characteristics of the prompt given, such as the number of lines and the position of the mutated part in the code. Furthermore, we assessed to what extent one-shot prompting (i.e., providing an example in the input prompt) could improve the MCT results compared to zero-shot prompting (i.e., providing no examples). 

Our case study results show that GPT-4 significantly outperforms GPT-3.5, regardless of mutation operators and programming languages, although even GPT-4 shows relative weaknesses in relational logic and Java programs.
However, the performance of GPT-3.5 could be dramatically improved using one-shot prompts instead of zero-shot prompts. 
The results also show that we can identify certain conditions (in terms of the mutation operators used and the number of tokens in the code) under which the LLMs result in correct or incorrect answers. 

The contributions of this paper are summarized as follows:
\begin{itemize}[$\bullet$]
    \item We present Mutation-based Consistency Testing (MCT), a new method for evaluating LLMs' code understanding capability (particularly subtle inconsistencies between code and its semantics).
    \item We present a case study demonstrating the applicability of MCT on GPT-3.5 and GPT-4.
    \item We investigate the strengths and weaknesses of GPT-3.5 and GPT-4 in terms of code semantic inconsistency identification, which can be used for future LLM-based SE studies.
    \item We will provide our replication package publicly available (see Section~\ref{sec:data}). 
\end{itemize}

The significance of our paper is as follows.
As LLM-based SE continue to gain more attention, evaluating them under a more systematic and generalized approach becomes indispensable, assessing not just their code generation but also their syntactic understanding and consistency. 
Using MCT, both researchers and practitioners can easily test their LLMs, thereby paving the way for more robust and reliable LLMs in the future.

The rest of the paper is organized as follows.
Section~\ref{sec:background} provides the essential basics, including LLMs and mutation analysis.
Section~\ref{sec:approach} presents our proposed method.
Section~\ref{sec:eval-design} presents our case study design.
Section~\ref{sec:results} shows case study results.
Section~\ref{sec:data} describes data availability.
Section~\ref{sec:related-work} reviews the relevant literature.
Section~\ref{sec:conclusion} concludes this paper and presents future research directions.

\section{Background}\label{sec:background}

\subsection{Large Language Models}\label{sec:background:llm}

Large Language Models (LLMs) are advanced Deep Learning (DL) systems designed to understand, generate, and interact with natural languages. Typically, LLMs are trained on vast datasets, encompassing a wide range of structured data, which enables them to grasp the nuances of language and context. Just as natural language sentences are structured data, code, as another kind of structured data is also used to train LLMs, enabling them to comprehend the syntax, semantics, and patterns in programs. This development marks a significant improvement in LLM-relevant research, extending the capabilities of LLMs beyond natural language to the realm of software development.

Generative Pre-trained Transformer (GPT) developed by OpenAI is one of the representative examples of successfully and widely used LLMs in recent years. The original GPT model~\cite{radford2018improving} laid the foundation for subsequent developments in the field. It was followed by GPT-2~\cite{radford2019language}, an advanced version with a larger number of parameters, renowned for its enhanced text generation capabilities. The series progressed with GPT-3~\cite{brown2020language}, a model enriched with 175 billion parameters, which demonstrated remarkable advancements in few-shot and zero-shot learning, significantly improving natural language understanding. The latest in this series, GPT-4~\cite{openai2023GPT-4}, further expanded the capabilities of its predecessors, offering more parameters and enhanced reliability, particularly in visual understanding and reasoning. Furthermore, GPT-based models, such as CodeGPT~\cite{lu2021codexglue} and Codex~\cite{chen2021evaluating} trained on extensive open-source code repositories, have significantly aided developers in various tasks~\cite{neurips2019codegen,fan2023large}. 

The advancement of programming-capable LLMs has been significantly accelerated by the introduction of code generation benchmarks, such as HumanEval~\cite{chen2021evaluating}, MBPP (MassiveBank of Python Problems)~\cite{austin2021program}, and CodeXGLUE~\cite{lu2021codexglue}. 
These benchmark datasets contain programming tasks (problems), covering multiple aspects such as language basics, algorithms, and mathematics. 
Each problem is accompanied by a canonical solution and a specific set of test inputs. 
The program generated by the LLM under test is then compared with the canonical solution by running both on the given test inputs. 
The LLM-generated program is considered correct only if its outputs match exactly with the outputs of the canonical solution for all the test inputs.
Figure~\ref{fig:example-problem} shows an example problem in HumanEval.

\begin{figure}
\begin{lstlisting}[language=Python, escapeinside={(*@}{@*)}]
(*@\textbf{Task ID:}@*)
Python 53

(*@\textbf{Problem:}@*)
def add(x: int, y: int):
    """
    Add two numbers x and y
    >>> add(2, 3)
    5
    """

(*@\textbf{Canonical Solution:}@*)
def add(x: int, y: int):
    return x + y

(*@\textbf{Test inputs:}@*)
assert add(0, 1) == 1
assert add(1, 0) == 1
...
\end{lstlisting}
\caption{An example problem, canonical solution, and test inputs in HumanEval~\cite{chen2021evaluating}}
\label{fig:example-problem}
\end{figure}

\subsection{Prompt Engineering}\label{sec:background:prompt}
Prompt engineering, a crucial element in the application of large language models (LLMs), has been extensively explored in recent research. \citet{wei2022chain} introduce the 'Chain-of-Thought' prompting method, where prompts are crafted to elicit a sequential reasoning process from LLMs, thereby enhancing their performance on complex tasks. Complementing this, \citet{shin2020autoprompt} present AutoPrompt, which uses gradient-based techniques for automatic prompt generation to effectively extract specific responses from language models. Additionally, \citet{brown2020language} explore the inherent capabilities of LLMs in their paper, particularly focusing on the emergent abilities as these models scale up in size and complexity.

Integration of language models with tree search algorithms, proposed by \citet{chen2021language}, enhances reasoning, acting, and planning capabilities in LLMs. In the context of programming, \citet{li2021octopack} investigate instruction tuning for large language models in their study, aiming to improve LLMs' performance in understanding and generating code. \citet{reynolds2021prompt} offer new insights into effective interaction with LLMs, moving beyond traditional few-shot learning approaches. These studies collectively underscore the dynamic nature of prompt engineering and its critical role in leveraging the full capabilities of LLMs in a variety of complex applications.

\subsection{Mutation Analysis}\label{sec:background:mutation}
Mutation analysis is a software testing method that seeks to evaluate the efficacy of a test suite by introducing small, systematic changes to the source code of a program. These intentional modifications, or mutations, result in a series of slightly altered versions of the program, i.e., a ``mutant''. If a test suite can detect these intentional faults, it's more likely to detect unintentional ones already present in the code. However, if it cannot detect intentional faults, it indicates potential weaknesses in the testing dataset. Figure~\ref{fig:mutation_demo} illustrates a simple mutant example. 

\begin{figure}
\centering
\begin{minipage}{0.45\linewidth}
\textbf{Original Program:}
\begin{lstlisting}[language=Python]
if (a and b) {
    c = 1
} else {
    c = 0
}
\end{lstlisting}
\end{minipage}
\hfill
\begin{minipage}{0.45\linewidth}
\textbf{Mutated Program:}
\begin{lstlisting}[language=Python]
if (a or b) {
    c = 1
} else {
    c = 0
}
\end{lstlisting}
\end{minipage}
\caption{A simplified mutant generation example. The logical operator `\texttt{and}' in the original program (left) has been modified to `\texttt{or}' in the mutant (right).}
\label{fig:mutation_demo}
\end{figure}

This mutation analysis concept has been explored and expanded in many papers. For instance, \cite{jia2011analysis} provide a comprehensive survey of the field, discussing its development, applications, and challenges. \cite{delgado2017systematic} offer a systematic review of mutation testing tools in their paper, comparing features and capabilities of various tools in this domain. \cite{papadakis2019mutation} provide an updated survey of advances in mutation testing, including recent developments. \cite{zhang2012predicting} discuss the use of mutation testing in improving fault localization, while \cite{wong2000mutation} discusses the challenges and potential of mutation testing at the turn of the century. Finally, \cite{just2014mutants} investigate the effectiveness of mutants as substitutes for real faults in software testing.

\section{Approach}\label{sec:approach}

We aim to systematically evaluate the code understanding capability of LLMs, particularly in terms of detecting subtle inconsistencies between code (written in programming languages) and its description (written in natural language).
To achieve this, we present Mutation-based Consistency Testing (MCT). 
The key idea behind the MCT approach is to deliberately inject artificial inconsistencies between code and its description using program mutation, which can simulate potential bugs that LLMs might encounter in real-world scenarios. 

Algorithm~\ref{alg:mtc} shows the pseudocode of our MCT approach. It takes as input the LLM under test $\mathcal{L}$, a set of consistent data $D = \{(\mathit{des}, \mathit{imp}), \dots \}$ where $(\mathit{des}, \mathit{imp})$ is a pair of a program description (i.e., $\mathit{des}$) and its corresponding implementation (i.e., $\mathit{imp}$), and a set of mutation operators $U$ applicable to the implementations in $D$.
The algorithm then returns an MCT score $s$ that ranges from 0 to 100, with larger values indicating better results.

Notice that $D$ is directly available from existing code generation benchmarks.
Also, obtaining $U$ is straightforward with existing mutation analysis studies across various programming languages.

\begin{algorithm}
\SetKwInOut{Input}{Input}
\SetKwInOut{Output}{Output}

\Input{LLM under Test $\mathcal{L}$, \\
    Set of Consistent Pairs $D = \{(\mathit{des}, \mathit{imp}), \dots \}$, \\
    Set of Mutation Operators $U$}
\Output{MCT Score $s$}

    Set of Passed Mutated Data $P \gets \emptyset$ \label{alg:mtc:p-init}\\
    Set of Failed Mutated Data $F \gets \emptyset$ \label{alg:mtc:f-init}\\

    \ForEach{Consistent Pair $(\mathit{des}, \mathit{imp}) \in D$}{\label{alg:mtc:for-d}
        \If{$\texttt{isCorrect}(\mathcal{L}, \mathit{des}, \mathit{imp})$}{\label{alg:mtc:org-corr}
            Set of Mutants $M \gets \texttt{genMutants}(\mathit{imp}, U)$ \label{alg:mtc:genM} \\

            \ForEach{Mutant $m\in M$}{\label{alg:mtc:for-m}
                \If{$\texttt{isCorrect}(\mathcal{L}, \mathit{desc}, m)$}{\label{alg:mtc:mut-corr}
                    $P \gets P \cup \{(\mathit{des}, m)\}$ \label{alg:mtc:cor}\\
                }
                \Else{
                    $F \gets F \cup \{(\mathit{des}, m)\}$ \label{alg:mtc:inc}\\
                }
            }
        }
    }
    MCT Score $s \gets \frac{|P|}{|P| + |F|} \times 100$ \label{alg:mtc:score}\\
    \textbf{return} $s$ \label{alg:mtc:return}
\caption{Mutation-based Consistency Testing (MCT) for LLMs}
\label{alg:mtc}
\end{algorithm}

The algorithm first initializes two sets: passed mutated data $P$ (line~\ref{alg:mtc:p-init}) and failed mutated data $F$ (line~\ref{alg:mtc:f-init}).
For each pair $(\mathit{des}, \mathit{imp}) \in D$ (line~\ref{alg:mtc:for-d}), the algorithm checks if $\mathcal{L}$ correctly answers ``consistent'' for the given $\mathit{des}$ and $\mathit{imp}$ (line~\ref{alg:mtc:org-corr}). Section~\ref{sec:prompt} details the prompt used to execute $\mathcal{L}$ using $\mathit{des}$ and $\mathit{imp}$.
The algorithm enters the main part of MCT (lines~\ref{alg:mtc:genM}--\ref{alg:mtc:inc}) only if $\mathcal{L}$ correctly answers. This ensures that the MCT score is computed purely based on the initially identified consistent pairs by $\mathcal{L}$.
In the main part of MCT, the algorithm first generates a set of mutants $M$ from the given code $\mathit{imp}$ using the mutation operators $U$. Section~\ref{sec:mutation} details the mutant generation process.
For each mutant $m\in M$ (line~\ref{alg:mtc:for-m}), the algorithm checks if $\mathcal{L}$ correctly answers ``inconsistent'' for the given $\mathit{des}$ and $m$ (line~\ref{alg:mtc:mut-corr}). If it is correct, then the pair of $\mathit{des}$ and $m$ is added to $P$ (line~\ref{alg:mtc:cor}); otherwise, the pair is added to $F$ (line~\ref{alg:mtc:inc}).
At the end, the algorithm ends by returning the MCT score $s = \frac{|P|}{|P|+|F|}\times 100$ (lines~\ref{alg:mtc:score}--\ref{alg:mtc:return}).

\subsection{Prompting}\label{sec:prompt}

Prompt engineering, as introduced in Section~\ref{sec:background:prompt}, is important as the design of prompts can significantly affect the output of LLMs. 
Inspired by existing methods used for testing LLMs on the HumanEval Dataset~\cite{chen2021evaluating}, we carefully crafted a prompt considering three essential aspects: (1) clear task definition, (2) structured input, and (3) explicit evaluation criteria. 
Figure~\ref{fig:prompt} shows the prompt template, with \{DESCRIPTION\} and \{CODE\} as placeholders for corresponding artifacts.

\begin{figure}
\begin{lstlisting}[escapeinside={(*@}{@*)}]
I'm presenting you with a program 
description and its corresponding code.
(*@\textbf{Description:} \textcolor{blue}{\{DESCRIPTION HERE\}}@*)
(*@\textbf{Code:} \textcolor{blue}{\{CODE HERE\}}@*)
Evaluate the code based on the description. 
Respond with:
1 - If the code functionally and syntactically matches the description completely without any minor inconsistency.
2 - Otherwise.
Please respond only with a single number: 
'1' or '2'.
\end{lstlisting}
\caption{Zero-shot prompt template with \{DESCRIPTION\} and \{CODE\} as placeholders for corresponding artifacts}
\label{fig:prompt}
\end{figure}

\begin{figure}
\begin{lstlisting}[language=Python, escapeinside={(*@}{@*)}]
(*@\textbf{Here is an example for you. When given this pair:}@*)

(*@\textbf{Description:}@*)
from typing import List\n\n\ndef has_close_elements(numbers: List[float], threshold: float) -> bool:
    """ Check if in given list of numbers, are any two numbers closer to each other than given threshold.    
    >>> has_close_elements([1.0, 2.0, 3.0], 0.5)
    False
    >>> has_close_elements([1.0, 2.8, 3.0, 4.0, 5.0, 2.0], 0.3)
    True
    """

(*@\textbf{Code:}@*)
for idx, elem in enumerate(numbers):
        for idx2, elem2 in enumerate(numbers):
            if idx != idx2:
                distance = abs(elem * elem2)
                if distance < threshold:
                    return True

    return False

(*@\textbf{The code did not match the description, the code should be:}@*)
for idx, elem in enumerate(numbers):
        for idx2, elem2 in enumerate(numbers):
            if idx != idx2:
                distance = abs(elem - elem2)
                if distance < threshold:
                    return True

    return False
    
(*@\textbf{So the correct answer is: 2}@*)
\end{lstlisting}
\caption{The additional input for one-shot prompts. This part is appended at the end of the zero-shot prompt template (Figure~\ref{fig:prompt}) to create one-shot prompts.}
\label{fig:one-shot-prompt}
\end{figure}

In the prompt template, we can see that 
(1) the task definition is provided clearly in the first part,
(2) the code and its description are presented separately in a structured manner, and
(3) the evaluation criteria are explicitly stated at the end. 

One-shot prompting is a common few-shot technique, which provides the LLM with a single example to guide its response. As discussed in Section~\ref{sec:background:llm}, it helps the LLM to better understand the context and the specific requirements of the task. 
Figure~\ref{fig:one-shot-prompt} shows the example part to be appended at the end of the zero-shot prompt (see Figure~\ref{fig:prompt}). 
The example serves as a reference for the LLM, illustrating how to analyze and respond to the task. By including this example, we could align the model's response mechanism more closely with the desired output, thereby enhancing the accuracy and relevance of its answers.
More experiments will follow in Section~\ref{sec:eval-design}.

\subsection{Mutant Generation}\label{sec:mutation}

To introduce inconsistencies between a description ($\mathit{des}$) and its implementation ($\mathit{imp}$), we systematically generate many mutants from $\mathit{imp}$. 
As discussed in Section~\ref{sec:background:mutation}, one can simply generate many subtle mutants by applying pre-defined mutation operators as much as possible. 

Although the process of generating mutations seems straightforward, it can sometimes produce mutants that are semantically equivalent to the original program. Since there is no inconsistency between a description and the equivalent mutant of its original program, we should avoid generating equivalent mutants. However, the equivalent mutant detection problem is known to be undecidable~\cite{Budd1982}. Nevertheless, this can be addressed by selecting mutation operators carefully to reduce the possibility of generating equivalent mutants~\cite{6823861}. Furthermore, we can leverage a rich set of equivalent mutant detection studies~\cite{6613487,7194639}.

Another potential issue is the sheer number of mutants. Oftentimes, the total number of mutants would be too much to deal with considering time and cost. Notice that mutant execution (using LLMs) is both time- and cost-intensive, not mutant generation. Therefore, one typical solution to this issue is a random sampling of mutants to execute after generation~\cite{Zhang10,7381815}. In other words, we can generate all possible mutants at first and then randomly sample some of them considering the available budget. 

\section{Case Study Design}\label{sec:eval-design}

In this section, we describe a case study to demonstrate the applicability and usefulness of our mutation-based consistency testing approach.
Although our approach is relatively intuitive and straightforward, it is unclear how useful the approach is in practice.
Therefore, we apply our approach to real-world LLMs with the aim of demonstrating the application of our approach.
To this end, we draw the following research questions:
\begin{enumerate}[\bf RQ1]
    \item How do different LLMs fare in terms of mutation-based consistency testing for different mutation operators? 
    \item How do different LLMs fare in terms of mutation-based consistency testing for different programming languages? 
    \item Can we explain the consistency testing results of LLMs in terms of input characteristics? 
    \item How do the consistency testing results of LLMs vary with zero-shot and one-shot promptings?
\end{enumerate}

RQ1 aims to evaluate the ability of different LLMs to determine the consistency between code and its description while considering the influence of different mutation operators applied to the code. 
It specifically focuses on two of the most commonly used LLMs, GPT-3.5~\cite{radford2018improving} and GPT-4~\cite{openai2023GPT-4}. 
The answer to RQ1 will help us understand when LLMs detect consistency between code and its description correctly.

Similar to RQ1, RQ2 also aims to test GPT-3.5 and GPT-4 using our approach. 
However, instead of focusing on mutation operators, RQ2 seeks to determine for which programming language the LLMs work well and for which they do not. 
The outcomes of RQ2 will provide insights into the testing performance of different programming languages.

In addition to RQ1 and RQ2, RQ3 takes a step further to explain under which conditions the LLMs correctly detect inconsistencies between code and its description. 
For instance, it might be the case that GPT-3.5 does not perform well when the code provided is quite lengthy. 
By examining these conditions in terms of input characteristics such as lines of code, programming language, mutation operator, and mutation position in the code, we can better understand the strengths and limitations of the LLMs.

RQ4 investigates the variation in the consistency testing results of LLMs when given no examples (zero-shot prompting) versus a single example (one-shot prompting). 
The findings from RQ4 will discover if minimal contextual information can greatly improve the model's ability to understand and analyze code, and will provide a deeper understanding of the capabilities and limitations of LLMs in processing complex programming tasks, and provide valuable insights for developers and researchers in optimizing the use of LLMs for code analysis.

\subsection{Datasets}\label{sec:dataset}

We used HumanEval-X~\cite{zheng2023codegeex}, an extension of the widely used open-source dataset HumanEval~\cite{chen2021evaluating}, as our dataset.

The HumanEval dataset is a hand-written evaluation set used to assess code generation ability for LLMs proposed by OpenAI.
It contains 164 Python programming tasks (see Figure~\ref{fig:example-problem} for an example task). 
Although HumanEval has been widely used to measure the (behavioral) correctness of LLM-generated code, it only contains Python programming tasks.
HumanEval-X extends HumanEval with six programming languages: Python, C++, Java, Go, JavaScript (JS), and Rust.
Given its expanded programming languages, HumanEval-X offers a more comprehensive benchmark, making it suitable for evaluating the latest and more advanced Code LLMs, including models like CodeGeeX~\cite{zheng2023codegeex} they proposed.

\subsection{Subject LLMs}\label{sec:subjects}

We chose GPT-3.5~\cite{radford2018improving} and GPT-4~\cite{openai2023GPT-4} as our test subjects.
Specifically, we used the stable version of both models, GPT-3.5-0613 and GPT-4-0613, using OpenAI APIs~\cite{openai_doc}. 

GPT-3.5~\cite{radford2018improving} is one of the most popular LLMs since it was first introduced in early 2022, particularly because ChatGPT is based on the GPT-3.5 series. 
In addition to its remarkable natural language processing capability, it achieves the HumanEval score of 47.1\%, meaning it generates correct solutions for 47.1\% of all programming tasks in HumanEval. 
This implies that GPT-3.5 is capable of code generation.

GPT-4~\cite{openai2023GPT-4}, a successor of GPT-3.5, is one of the most powerful LLMs at the time of writing this paper. 
Specifically, it achieves the HumanEval score of 67.0\%, which is human-level performance. 
By comparing and contrasting GPT-3.5 and GPT-4 in our case study, we can analyze the two subsequent models.

In our preliminary experiments, we discovered two issues with both GPT-3.5 and GPT-4 models. 
Firstly, they were nondeterministic by default, meaning that when prompted with the same input, they would return different outputs across multiple executions. 
Secondly, they often returned more than one output with unnecessary details, even though the prompt explicitly requested a single number (see Section~\ref{sec:prompt} for more details about our prompting). 

To address the issues, we adjusted two parameters: \texttt{temperature} and \texttt{max\_tokens}. 
According to OpenAI API documentation~\cite{openai_doc}, lower values for temperature, such as 0.2, result in more consistent outputs, while higher values generate more diverse and creative results, such as 1.0. 
The value of \texttt{max\_tokens} specifies the maximum number of tokens in the model's response. It is recommended to set its value as close to the expected response size as possible.
Therefore, to make the models deterministic, we set \texttt{temperature} to 0.2 throughout our experiments. Additionally, since we only needed a single number as a response, we set the \texttt{max\_tokens} to 1.

\subsection{Mutation Operators}
As explained in Section~\ref{sec:mutation}, mutation operators are essential to systematically introduce subtle changes to the code.
For our case study, we selected mutation operators according to the following criteria:
(1) widely used in the literature~\cite{just2014mutants,papadakis2019mutation,delgado2017systematic}, and
(2) applicable to all six programming languages (i.e., Python, C++, Java, Go, JavaScript, and Rust). 
As a result, we chose four mutation operators: 
Arithmetic Operator Replacement (AOR),
Relational Operator Replacement (ROR),
Literal Value Replacement (LVR), and
Statement Deletion (STD).
Table~\ref{tab:mutation_operators} summarizes the mutation operators with simple examples. 
Notice that each mutation operator exploits different aspects of the code, allowing us to investigate the strengths and weaknesses of the LLMs in consistency testing.

\begin{table}
    \centering
    \caption{Summary of Mutation Operators}
    \begin{tabular}{lll}
    \toprule
    \textbf{Name} & \textbf{Description} & \textbf{Example} \\
    \midrule
    AOR & Arithmetic Operator Replacement & a \textcolor{blue}{$+$} b $\to$ a \textcolor{red}{$-$} b \\
    LVR & Literal Value Replacement & \textcolor{blue}{10} $\to$ \textcolor{red}{9} \\
    ROR & Relational Operator Replacement & a \textcolor{blue}{$<$} b $\to$ a \textcolor{red}{$>=$} b \\
    STD & Statement Deletion & remove one line \\
    \bottomrule
    \end{tabular}
    \label{tab:mutation_operators}
\end{table}

We initially generated a total of 38018 mutants by applying all possible mutation operators. 
However, considering the time and cost involved in executing GPT-3.5 and GPT-4, we randomly selected 100 mutants for each programming language and mutation operator. 
Therefore, our case study considered a total of 2400 mutants (i.e., 100 randomly selected mutants $\times$ 4 mutation operators $\times$ 6 programming languages). 
Notice that we used the same mutants for both GPT-3.5 and GPT-4 to ensure a fair comparison.

\subsection{Methodologies}\label{sec:methodology}
This subsection describes methodologies to address our research questions. 

\subsubsection{(RQ1, RQ2) Mutation Operators and Programming Languages}
To answer RQ1 and RQ2, we repeated Algorithm~\ref{alg:mtc} for the two LLMs using the same HumanEval-X dataset and the four mutation operators. 
By recording the programming languages and mutation operators used for each LLM execution, we could compute the MCT score for each programming language and mutation operator.

\subsubsection{(RQ3) Input Characteristic Analysis}\label{sec:meth-tree}
To answer RQ3, we used the data collected for RQ1 and RQ2 and built decision trees to infer how the consistency testing results (i.e., correct or incorrect) relate to input characteristics (e.g., lines of code provided in the prompt). We used decision trees because they are easy to interpret and useful in identifying important features. More important features appear closer to the root, and less important ones may be even removed during pruning~\cite{weka}. 

Since our target variable (i.e., the consistency testing result) is binary, we used a classification tree. Specifically, we built a classification tree to predict the consistency testing result of a given input prompt based on the following input characteristics as features:
\begin{itemize}[$\bullet$]
    \item Model (nominal): GPT-3.5 or GPT-4
    \item Mutation operator (nominal): AOR, ROR, LVR, or STD
    \item Programming language (nominal): Python, C++, Java, Go, JS, or Rust
    \item Line (numeric): Total lines of code
    \item Token (numeric): Total number of tokens in the code
    \item Mutation position (numeric): Position (line number) of mutation in the code
\end{itemize}
In addition to the nominal features (i.e., model, mutation operator, and programming language), we added the numeric features (i.e., lines, tokens, and mutation position) as the static characteristics of the input code. 

To evaluate the predictive accuracy of the generated classification tree, we measured the percentage of correctly classified instances using 10-fold cross validation on the 4800 data instances (i.e., 2400 mutants for each of the two LLMs).
Based on our preliminary experiments, we set the minimum number of observations per leaf node to 150 to balance the interpretability and accuracy of the tree.
We used the default values provided by Weka for the other parameters.

\subsubsection{(RQ4) One-Shot Prompting}
To answer RQ4, we used the zero-shot and one-shot prompts based on the template shown in Figure~\ref{fig:prompt}. Due to the high execution cost of GPT-4, we used only GPT-3.5. 
\section{Case Study Results}\label{sec:results}

\subsection{RQ1: Impact of Mutation Operators}
Table~\ref{table:operators} shows how the performance of both GPT-4 and GPT-3.5 varies on different mutation operators. The bottom row displays the difference in MCT scores between the two models.
For example, for the same set of mutants generated by applying the AOR mutation operator, GPT-3.5 and GPT-4 achieve MCT scores of 38.7 and 90.3, respectively, and as a result, the difference is 51.7. 

\begin{table}
\caption{MCT Score for Different Mutation Operators}\label{table:operators}
\centering
\begin{tabular}{lrrrrr}
\toprule
Model & AOR & LVR & ROR & STD & Avg \\
\midrule
GPT-3.5 & 38.7 & 20.2 & 25.7& 51.3 & 34.0 \\
GPT-4 & 90.3 & 83.5 & 81.5 & 85.2 & 85.1 \\
\midrule
Diff & 51.6 & 63.3 & 55.8 & 33.9 & 51.1 \\
\bottomrule
\end{tabular}
\end{table}

Overall, it is clear that GPT-4 outperforms GPT-3.5, regardless of the mutation operators. This indicates that GPT-3.5's ability to understand code and its descriptions as well as to identify subtle inconsistencies between them is considerably inferior to that of GPT-4. This is likely due to the vast amount of training data and trainable parameters in GPT-4, in comparison to GPT-3.5.

As for GPT-3.5, the MTC scores range from 20.7 (LVR) to 51.3 (STD), indicating its poor ability to identify subtle inconsistencies between code and its description. Specifically, the lowest score of 20.7 for LVR implies that GPT-3.5 suffers from identifying minor changes in literal values, which is a crucial aspect in many SE tasks, such as debugging. It is perhaps intuitive that STD has the highest score of 51.3, given that deleting a statement is relatively easier to detect than other changes in AOR, LVR and ROR. However, a score of 51.3 means that almost half of the deleted statements are not accurately detected by GPT-3.5, which is surprising. 

As for GPT-4, the MCT scores for all mutation operators are above 80, indicating its decent ability to identify subtle inconsistencies between code and its description. On the one hand, it achieves the highest score of 90.3 for AOR, which highlights its proficiency in reasoning arithmetic operators. On the other hand, the lowest score of 81.5 is for ROR, which suggests that GPT-4 has relatively less accuracy in understanding relational logic.

\begin{tcolorbox}
The answer to RQ1 is that, regardless of mutation operators used, GPT-4 significantly outperforms GPT-3.5 in terms of the MCT scores. However, GPT-4 shows the lowest score of 81.5 for ROR, indicating the relative weakness of dealing with relational logic. 
\end{tcolorbox}

\subsection{RQ2: Impact of Programming Language}
Table~\ref{table:languages} shows the MCT scores of both GPT-3.5 and GPT-4 for different programming languages. The results indicate a clear pattern of GPT-4 outperforming GPT-3.5 across all languages. The bottom row of the table highlights the difference in MCT scores between the two models, showcasing the advancement in GPT-4's capabilities.

\begin{table}[ht]
\caption{MCT Score for Different Programming Languages}\label{table:languages}
\centering
\begin{tabular}{lrrrrrrr}
\toprule
Model & Python & C++ & Go & Java & JS & Rust & Avg \\
\midrule
GPT-3.5 & 29.0 & 28.8 & 30.5 & 27.5 & 44.8 & 43.3 & 34.0 \\
GPT-4 & 89.3 & 89.0 & 81.3 & 77.8 & 86.0 & 87.5 & 85.1 \\
\midrule
Diff & 60.3 & 60.2 & 50.8 & 50.3 & 41.2 & 44.2 & 51.1 \\
\bottomrule
\end{tabular}
\end{table}

On the one hand, GPT-3.5 shows a relatively consistent but low MCT score across all programming languages, with its highest performance in JavaScript (44.8) and lowest in Java (27.5). This is somewhat consistent with \cite{abdulkareem2021evaluating} who ranked the complexity of programming languages from least to most complex as Python, C++, JavaScript, and Java. 
This suggests that while GPT-3.5 has a general understanding of programming languages, it struggles with the complexities of specific languages like Java.

On the other hand, GPT-4 demonstrates a remarkable improvement in all languages, particularly excelling in Python (89.3) and C++ (89.0). Its lowest score is in Java (77.8), but it still significantly outperforms GPT-3.5. This indicates GPT-4's advanced ability to understand and process various programming languages, likely due to its more diverse and extensive training data.

The analysis also reveals specific strengths and weaknesses of each model. GPT-3.5's lower scores across the board suggest difficulties in handling complex syntax and semantics, especially in languages like C++ and Rust. In contrast, GPT-4's high scores in these languages indicate a strong grasp of complex data types and memory management concepts.

Also, interestingly, the trend of both models on the same programming language is different. Java, Python, and C++ are the least-performed languages on GPT-3.5, yet only Java is still the weakness of GPT-4 and Python and C++ become the best-performed languages in GPT-4. This implies the possible change in the training dataset and significant structural change in the GPT-4, leading to a huge improvement in performance.

\begin{tcolorbox}
The answer to RQ2 is that GPT-4 significantly outperforms GPT-3.5 in understanding and processing various programming languages. 
GPT-4's performance gain is particularly notable in languages with simple syntax and semantics, such as Python and C++.
\end{tcolorbox}

\subsection{RQ3: Explaining Test Results}

\begin{figure}[ht]
\centering
\includegraphics[width=\linewidth]{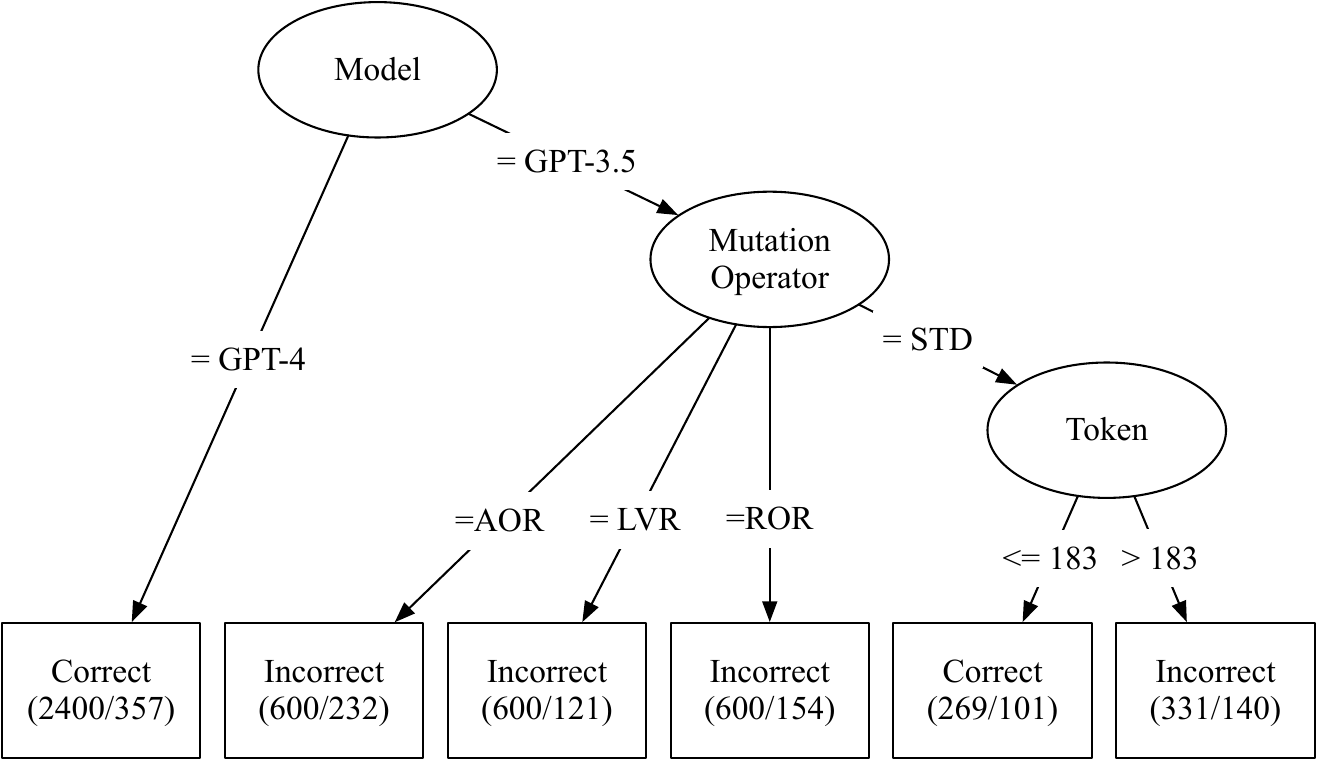}
\caption{Model Decision Tree}
\label{fig:model_diagram}
\end{figure}

Figure~\ref{fig:model_diagram} shows the classification tree built following the methodology described in Section~\ref{sec:meth-tree}. The decision tree consists of non-leaf nodes (circles) and leaf nodes (squares). Each non-leaf node corresponds to a significant feature (input characteristic) that impacts the condition, while each leaf node represents the predicted MCT outcome (either Correct or Incorrect) for the condition associated with the path from the root to the leaf. In each leaf node, the first value in the parenthesis represents the total number of data instances (inconsistent pairs) from the training set that fall into that leaf and the second value represents the number of incorrectly classified instances. For example, the left-most leaf node indicates that the MCT result is correct when the model is GPT-4, which has 357 misclassified instances out of 2400. 

Overall, the non-leaf nodes in the tree clearly show that the factors that have the greatest influence on the performance of the model are `model', `mutation operator', and `length', in that order. It is worth noting that for GPT-4, `mutation operator' and `length' are not as important since the model works well independently from these factors, which is consistent with the results of our RQ1 and RQ2. However, for GPT-3.5, the tree reveals that the performance of the model varies significantly depending on the mutation operator used. Furthermore, we can see that GPT-3.5 is incorrect for longer code, longer than 183 in our dataset. Interestingly, the tree considers `length` not `line` significant, meaning that the number of tokens matters more than the number of lines. It seems intuitive because having more lines in the prompt would only result in having more new-line characters.

\begin{tcolorbox}
The answer to RQ3 is that we can identify conditions under which GPT-3.5 and GPT-4 work well or not in terms of different input characteristics. Although GPT-4 is generally correct, the correctness of GPT-3.5 depends on the mutation operators used and the number of tokens in the code. 
\end{tcolorbox}

\subsection{RQ4: Impact of Prompting}
Table~\ref{table:prompt-engineering} illustrates the impact of prompt engineering, specifically zero-shot and one-shot prompts, on the MCT scores of the GPT-3.5 model across various programming languages and mutation operators. The column \textit{Impr} in the table quantifies the improvement of one-shot prompts over zero-shot prompts, highlighting the effectiveness of this technique.

\begin{table}[ht]
\caption{MCT Score for Zero-Shot and One-Shot Prompts}
\label{table:prompt-engineering}
\centering
\begin{tabular}{llrrr}
\toprule
Language & MO & Zero-Shot & One-Shot & Impr \\ 
\midrule
Python & AOR & 42 & 93 & 51 \\
       & LVR & 25 & 89 & 64 \\
       & ROR & 12 & 70 & 58 \\
       & STD & 35 & 89 & 54 \\
\midrule
C++    & AOR & 23 & 80 & 57 \\
       & LVR & 13 & 79 & 66 \\
       & ROR & 21 & 85 & 64 \\
       & STD & 40 & 81 & 41 \\
\midrule
Go     & AOR & 32 & 87 & 55 \\
       & LVR & 11 & 79 & 68 \\
       & ROR & 33 & 85 & 52 \\
       & STD & 38 & 81 & 43 \\
\midrule
Java   & AOR & 34 & 84 & 50 \\
       & LVR & 13 & 61 & 48 \\
       & ROR & 15 & 63 & 48 \\
       & STD & 52 & 78 & 26 \\
\midrule
JavaScript & AOR & 45 & 93 & 48 \\
           & LVR & 29 & 90 & 61 \\
           & ROR & 62 & 94 & 32 \\
           & STD & 55 & 90 & 35 \\
\midrule
Rust   & AOR & 56 & 100 & 44 \\
       & LVR & 33 & 100 & 67 \\
       & ROR & 37 & 99 & 62 \\
       & STD & 35 & 95 & 60 \\
\midrule
Avg & - & 32.96 & 85.21 & 52.25 \\
\bottomrule
\end{tabular}
\end{table}

The overall pattern emerging from the table is a significant enhancement in the model's performance when employing one-shot prompts. This improvement is evident across all programming languages and mutation operators. For example, in Python, the MCT score for AOR mutations increases dramatically from 42 to 93 when shifting from zero-shot to one-shot prompting. Similarly, in Rust, the MCT score for AOR mutations reaches a perfect 100 with one-shot prompting, up from 56 in the zero-shot scenario.

This consistent improvement across different languages and mutation types indicates the profound impact that even a single, well-crafted example (e.g., Figure~\ref{fig:one-shot-prompt}) can have on the model's understanding and response accuracy. The most notable gains are observed in cases where the zero-shot approach yielded lower scores, suggesting that prompt engineering is particularly effective in addressing specific weaknesses in the LLM's performance.

The results underscore the critical role of prompt engineering in maximizing the potential of LLMs like GPT-3.5. The substantial increase in accuracy with one-shot prompts is a key insight for practical applications, especially in software development tasks such as automated code analysis and debugging. It demonstrates that providing minimal contextual guidance through a single example can significantly boost an LLM's performance, making it a more effective tool for nuanced and context-sensitive tasks in software engineering and beyond.

\begin{tcolorbox}
The answer to RQ4 is that simply using one-shot prompts instead of zero-shot prompts significantly enhances the performance of GPT-3.5 in terms of MCT scores. This highlights the effectiveness of providing even minimal contextual guidance to LLMs in improving their accuracy and adaptability in complex tasks.
\end{tcolorbox}

\subsection{Threats to Validity}

Bugs in our implementation of the experiment scripts could be a potential threat to the validity of the case study results. To mitigate this, we performed multiple code reviews for the core implementations. Furthermore, we manually verified the input and output of LLMs using various benchmark problems to ensure that everything was automated as expected. To improve transparency, we plan to make the implementation publicly available under an open-source license (see Section~\ref{sec:data}).

The four mutation operators used in the case study could be another potential threat. To address this issue, we carefully selected the mutation operators from widely used and well-accepted mutation testing tools and papers. These operators are derived from real-world errors commonly found in code issues and are applicable to most of our target programming languages. 
However, we want to note that our approach is independent of a specific set of mutation operators, and the selection of mutation operators is not the main contribution of this work.
In practice, more mutation operators applicable to a specific programming language can be used. 
Nevertheless, more studies on using various mutation operators are encouraged. 

A potential factor that may affect our results is whether the model accurately understands the natural language description in the input. 
However, according to the existing research on text understanding~\cite{devlin2018bert, rogers2020primer} and benchmarks~\cite{wang2018glue, liu2019roberta}, the error caused by text understanding is minimal. Nevertheless, more studies are needed to explore the nuances of language model comprehension in different contexts.

\section{Data Availability}\label{sec:data}

The replication package of our paper, including the implementation of the MCT method, the generated mutants, and the execution results of GPT-3.5 and GPT-4, is available at \url{https://figshare.com/s/70b466c68d5d7542dcd0}.
We plan to make it publicly available with an open-source license upon acceptance. 
\section{Related Work}\label{sec:related-work}

The research landscape of LLMs has expanded rapidly, with various studies focusing on their capabilities, limitations, and potential applications. This section provides an overview of the most relevant literature in the domains of LLMs testing on code-related tasks.

We found that the literature on testing LLM's programming capability can be categorized into three parts: code generation (including code completion), code understanding (including code summarization and code AST generation).

\subsection{Code Generation} 
This is a hot topic and has the most usage in real-world applications. There are several benchmarks and research papers to evaluate the code generation ability of LLMs. 

Two major benchmarks are HumanEval~\cite{chen2021evaluating}, MBPP~\cite{austin2021program}.
HumanEval focuses on assessing code synthesis models with a variety of programming challenges, each accompanied by a function signature, description, and test cases. Complementing this, MBPP is a collection of mostly basic Python problems, aimed at evaluating models on fundamental programming tasks and basic algorithms. 

Recently, HumanEval+~\cite{zheng2023codegeex} enhances HumanEval, featuring a broader range of more complex programming problems. HumanEval-X~\cite{zheng2023codegeex} extends HumanEval to include more programming languages. 
This progression of datasets signifies the evolving landscape of benchmarks designed to rigorously test and refine the capabilities of LLMs for programming.

There is also much research focusing on discovering the code generation ability in detail. 
\citet{troshin2022probing} proposed a diagnostic tool to test LLMs on different code-related tasks and examine how various aspects of the model, such as pre-training, model size, and fine-tuning, affect the results.
\citet{tian2023chatgpt} evaluated GPT-3.5 on its ability in code generation, program repair and code summarization and found that unrelated prompts have defects in the accuracy of the model.

Although assessing the code generation ability of LLMs has been widely studied as such, these studies focus on evaluating the output of LLMs in code-related tasks and provide no evaluation of how well can LLMs understand code semantics.

\subsection{Code Understanding} 
This line of work is more about understanding the overall code structure instead of code semantics.
\citet{shen2022benchmarking} proposed a benchmark to measure the performance of code syntax understanding and found out most LLMs have difficulty recognizing the syntactic relations in programs.
Recently, \citet{ma2023scope} further introduced several code LLM evaluation techniques with fine-grained categories of code understanding: code syntax understanding, code static behaviour understanding, and code dynamic behaviour understanding. Although these categories cover a wide range of code syntax and structural aspects, there is no evidence to suggest that LLMs can correctly identify the semantics of codes.

Notably, \citet{ma2023scope} introduced an evaluation method for code dynamic behavior understanding using mutation analysis. Specifically, they performed an equivalent mutant detection test on ChatGPT, where they prompted ChatGPT to determine whether a given mutant was equivalent to the original code. 
This is one of the most close studies to our work. 
However, they gave the LLM under test both the original program and the mutant, whereas we gave the LLM the code and its natural language description. As a result, we can test whether the LLM can correctly detect subtle inconsistencies between the code and its semantics written in natural language. 

\subsection{Summary}
To summarize, none of the existing studies focus on whether LLMs are capable of understanding code semantics, as well as minor changes or errors in the code. Evaluating both natural language and programming language for their consistency at the same time is our originality. By merging concepts of mutation analysis from software testing and code generation benchmarks for LLMs, our MCT method can assess the detailed performance of LLMs in terms of code semantic understanding on different aspects of code.

\section{Conclusion}\label{sec:conclusion}

To test the code understanding ability of LLMs, especially semantics and minor inconsistency of the code, we proposed Mutation-based Consistency Testing (MCT) and conducted a case study on GPT-3.5 and GPT-3.5. Our method is capable of finding the detailed performance of each model on various programming languages and mutation operators and identifying the weaknesses of each model in various aspects. 
Also, by further analyzing the MCT results based on decision tree analysis, we discovered the sensitivity of GPT-3.5 on input token length. Lastly, by comparing zero-shot and one-shot prompt results on GPT-3.5, we found that appending one simple example at the end of the prompt can improve the performance of GPT-3.5 greatly, even outperforming GPT-4.

For future work, we plan to conduct a more comprehensive evaluation encompassing a wider array of LLMs and mutation operators. We also plan to create a benchmark that enables both researchers and practitioners to effectively evaluate and compare various models across multiple dimensions. Additionally, one could use MCT results to improve the performance of LLMs. For example, by integrating MCT into the training process, it would be possible to enhance the accuracy of the models, thereby advancing their capabilities in real-world programming applications. 

\bibliographystyle{ACM-Reference-Format}
\bibliography{references}

\end{document}